\documentstyle[12pt]{article}
\begin{document}
\rightline{RUB-TPII-21/98}
\rightline{hep-ph/9901425}
\vspace{.3cm}
\begin{center}
\begin{large}
{\bf Nucleon matrix elements of twist--3 and 4 operators from the 
instanton vacuum$^\dagger$}
\end{large}
\\[2.5cm]
{\bf J.\ Balla}$^{\rm a}$, {\bf M.V.\ Polyakov}$^{\rm a, b}$
{\bf and C. Weiss}$^{\rm a}$  
\\[1.cm]
$^{a}${\em Institut f\"ur Theoretische Physik II,
Ruhr--Universit\"at Bochum, \\ D--44780 Bochum, Germany} 
\\
$^{b}${\em Petersburg Nuclear Physics Institute, Gatchina, \\
St.Petersburg 188350, Russia} 
\end{center}
\vspace{1.5cm}
\begin{abstract}
\noindent
The spin--dependent twist--3 and 4 nucleon matrix elements
$d^{(2)}$ and $f^{(2)}$ describing power corrections to the Bjorken and
Ellis--Jaffe sum rules are computed in the instanton vacuum. A systematic
expansion in the small packing fraction of the instanton medium, 
$\bar\rho / \bar R \ll 1$, is performed.  We find that the twist--3 matrix
element $d^{(2)}$ is suppressed [of order $(\bar\rho / \bar R)^4$], while
the twist--4 matrix element $f^{(2)}$ is of order unity.  Numerically, 
$d^{(2)} \ll f^{(2)}$.  The small value of $d^{(2)} \sim 10^{-3}$ obtained
from instantons is consistent with the recent E143 measurements of the
structure function $g_2$, where $d^{(2)}$ enters at the same level as the
twist--2 contribution.
\end{abstract}
\vfill
\rule{5cm}{.15mm} \\
{\footnotesize $\dagger$ Oral contribution to the 8th International 
Conference on the Structure of Baryons (Baryons '98), Bonn,
Sep.\ 22--26, 1998; to appear in the Proceedings.}
\newpage
In QCD, the quantitative description of deep--inelastic scattering and
other related processes comes down to a study of matrix elements of certain
composite operators in the nucleon state. The well--known parton
distributions, which parametrize the non--power suppressed part of the
structure functions, are given by matrix elements of operators of twist
2. In addition, there is a large number of observables related to operators
of non--leading twist ($>2$). One example are power corrections to the
Bjorken sum rule \cite{SV82,EhrMS94}.  To order $1/Q^2$:
\begin{eqnarray}
\int_0^1 dx \, g_1^{p - n} (x, Q^2 )
&=& \frac{1}{6} a^{(0)}_{\mbox{\scriptsize NS}}
\; + \; \frac{M_N^2}{27 Q^2} \left( a^{(2)}_{\mbox{\scriptsize NS}} 
\; + \; 4 d^{(2)}_{\mbox{\scriptsize NS}} 
\; + \; 4 f^{(2)}_{\mbox{\scriptsize NS}}
\right) ,
\label{g1}
\end{eqnarray}
where $a^{(0)}, a^{(2)}$ denote the matrix elements of twist--2 operators,
with $a^{(0)}_{\mbox{\scriptsize NS}} = (g_A )_{\mbox{\scriptsize NS}}$,
while $d^{(2)}$ and $f^{(2)}$ parametrize, respectively, the proton matrix
element of the twist--3 spin--2 and the twist--4 spin--1 operators:
\begin{eqnarray}
\lefteqn{ \langle P S | O^{\alpha\beta\gamma}_{\mbox{\scriptsize NS}} +
O^{\beta\alpha\gamma}_{\mbox{\scriptsize NS}}
| P S \rangle - \mbox{traces}} && \nonumber \\
&=& 2 M_N \; d^{(2)}_{\mbox{\scriptsize NS}}
\; \left[ 2 P^\alpha P^\beta S^\gamma
- P^\gamma P^\beta S^\alpha - P^\alpha P^\gamma S^\beta
+ (\alpha \leftrightarrow \beta ) - \mbox{traces} \right] , \;\;\;\;
\label{d2_def}
\end{eqnarray}
\begin{equation}
\langle P S | 
O^{\alpha\beta\alpha}_{\mbox{\scriptsize NS}}
| P S \rangle 
\;\; = \;\; 2 M_N^3 \; f^{(2)}_{\mbox{\scriptsize NS}}
\; S^\beta ,
\label{f2_def}
\end{equation}
\begin{equation}
O^{\alpha\beta\gamma}_{\mbox{\scriptsize NS}} (x) 
\;\; = \;\; g \, \bar\psi (x) \, 
\tau^3 \, \gamma^\alpha \, \widetilde{G}^{\beta\gamma} (x) \, \psi (x) .
\label{O}
\end{equation}
Knowledge of the values of $d^{(2)}$ and $f^{(2)}$ would allow one to
extend the QCD analysis of the Bjorken and Ellis--Jaffe sum rules to low
values of $Q^2$, where data with good statistics is available, which could
be of interest {\it e.g.}\ for an accurate determination of $\alpha_S$.
Similar expressions can be written for the flavor--singlet part of $g_1$
(Ellis--Jaffe SR), as well as for the GLS sum rule for $F_3$, see
Refs.\cite{SV82,BPW97}.
\par
It should be stressed that the parameters $d^{(2)}, f^{(2)}$, and the like,
are universal characteristics of the nucleon, whose significance extends
far beyond power corrections to $g_1$. For instance, the twist--3 matrix
element $d^{(2)}$ appears at leading twist level ({\it i.e.} not power
suppressed relative to twist--2) in the third moment of the structure
function $g_2$ \cite{SV82,EhrMS94}:
\begin{equation} 
\int_0^1 dx \, x^2 g_2^{p - n} (x, Q^2 ) 
\;\; = \;\; -\frac{1}{9} a^{(2)}_{\mbox{\scriptsize NS}}
\; + \; \frac{1}{9} d^{(2)}_{\mbox{\scriptsize NS}} 
\; + \; O\left(\frac{1}{Q^2}\right) 
\label{g2}
\end{equation}
Also, the twist--4 matrix element $f^{(2)}$ can be related, via the
heavy--quark expansion, to the charm contribution to $g_1$ \cite{PST98}.
Thus, these matrix elements can in principle be determined by combining
results of different measurements.  In order to compute them from first
principles, one needs a theory of the nonperturbative effects giving rise
to quark--gluon correlations in the nucleon.
\par
There is by now overwhelming evidence --- both from phenomenology as well
as from lattice simulations \cite{lattice} --- for an important role of
instanton--type vacuum fluctuations in determining the properties of the
low--energy hadronic world \cite{SchSh96}. The so--called instanton vacuum
approximates the ground state of Yang--Mills theory by a medium of
instantons and antiinstantons, with quantum fluctuations about them, which
is stabilized by instanton interactions \cite{DP84}. The coupling constant
is fixed at a scale of the order of the inverse average size of instantons
in the medium, $\bar\rho^{-1} \simeq 600\, {\rm MeV}$. The most important
property of this picture is the diluteness of the instanton medium, meaning
that the ratio of the average size of instantons in the medium to their
average distance, $\bar R$, is small: 
$\bar\rho / \bar R \simeq 1/3$ \cite{SchSh96,DP84}. The existence of this
small parameter makes possible a systematic analysis of non-perturbative
effects in this approach.
\par
In particular, the instanton vacuum explains the dynamical breaking of
chiral symmetry \cite{DP86,D96_Varenna}.  It happens due to the
delocalization of the fermionic zero modes associated with the individual
(anti--) instantons in the medium. Using the $1/N_c$--expansion one derives
from the instanton vacuum an effective low--energy theory, whose degrees of
freedom are pions (Goldstone bosons) and massive ``constituent'' quarks. It
is described by the effective action \cite{DP86}
\begin{equation}
S_{\rm eff} \;\; = \;\; \int d^4 x \;
\bar\psi (x) \left[ 
i \gamma^\mu \partial_\mu \; - \; 
M F(\stackrel{\leftarrow}{\partial} ) e^{i \gamma_5 \tau^a \pi^a (x)}
F(\stackrel{\rightarrow}{\partial} ) \right] \psi (x) .
\label{S}
\end{equation}
Here, $M$ is the dynamical quark mass generated by the spontaneous breaking
of chiral symmetry; parametrically
\begin{eqnarray}
M\bar\rho &\sim & \left(\frac{\bar\rho}{\bar R}\right)^2 ,
\end{eqnarray}
and $F(k)$ is a form factor proportional to the wave function of the
instanton zero mode, which drops to zero for momenta of order 
$k \sim \bar\rho^{-1}$.  Mesonic correlation functions, computed either
with the effective action, Eq.(\ref{S}), using the $1/N_c$--expansion
\cite{DP86}, or by more elaborate numerical simulations \cite{SchSh96},
show excellent agreement with phenomenology. The nucleon can be described,
within the $1/N_c$--expansion, as a chiral soliton of the effective theory
\cite{DPP88}. This field--theoretic description of the nucleon gives a good
account not only of its static properties, but also of the twist--2 parton
distributions \cite{DPPPW96}.
\par
Given the general success of the instanton vacuum in describing hadronic
properties, it is natural to apply this approach to the calculation of
matrix elements of higher--twist operators such as Eq.(\ref{d2_def}) and
(\ref{f2_def}). A general method for computing matrix elements of
quark--gluon operators in the instanton vacuum has been developed in
Ref.\cite{DPW96}. Due to the diluteness of the instanton medium the
dominant contribution to the matrix element is obtained by substituting for
the gluon operator in Eqs.(\ref{d2_def}) and (\ref{f2_def}) the classical
gluon field of a {\it single} (anti--) instanton. The interaction of the
instanton with the fermions through the zero modes then allows one to
replace the original QCD operator (normalized at $\mu = \bar\rho^{-1}$) by
an effective operator, expressed in terms of the quark and pion fields of
the effective theory.  It was shown in Ref.\cite{BPW97} that identities
following from the QCD equations of motion are fully preserved in this
approach.  For the operator Eq.(\ref{O}) the effective operator, after
integration over the instanton coordinates, takes the form:
\begin{eqnarray}
\mbox{``$O$''}_{\alpha\beta\gamma} (x) &=&
\bar\psi (x) \tau^3 \frac{\lambda^a}{2} \gamma_\alpha \psi (x)
\times
\frac{(-M)}{N_c} \int d^4 z \; f_{\beta\gamma , \mu\nu} 
(x - z) 
\nonumber \\
&& \times
\left[\bar\psi (z) \frac{\lambda^a}{2} \gamma_5 \sigma_{\mu\nu}
F(\stackrel{\leftarrow}{\partial} ) e^{i \gamma_5 \tau^a \pi^a (z)}
F(\stackrel{\rightarrow}{\partial} ) \psi (z) \right] ,
\label{O_eff}
\end{eqnarray}
where the function $f_{\beta\gamma , \mu\nu} (x)$ describes the (anti--)
selfdual field of the (anti--) instanton in singular gauge,
\begin{eqnarray}
G_{\beta\gamma}^a (x)_{I (\bar I)} &=&
\pm \widetilde{G}_{\beta\gamma}^a (x)_{I (\bar I)} 
\;\; = \;\; \frac{1}{g} f_{\beta\gamma , \mu\nu} (x) 
\left\{ \begin{array}{c} \bar\eta^{a}_{\mu\nu} \\[1ex] \eta^{a}_{\mu\nu}
\end{array} \right. ,
\\
f_{\beta\gamma , \mu\nu} (x) &=&
\frac{8  \rho^2 }{(x^2 + \rho^2 )^2}
\left( \frac{x_\mu x_\beta}{x^2} \delta_{\gamma\nu} 
+ \frac{x_\nu x_\gamma}{x^2} \delta_{\mu\beta} 
- \frac{1}{2} \delta_{\mu\beta} \delta_{\gamma\nu} \right) .
\end{eqnarray}
In Eq.(\ref{O_eff}) the gluon field of the QCD operator has been replaced
by a chirally--odd instanton--induced quark vertex proportional to the
dynamical quark mass, $M$. Note that, as in the effective Lagrangian,
Eq.(\ref{S}), chiral invariance is preserved due the presence of the pion
field.
\par
The matrix element of the effective operator, Eq.(\ref{O_eff}), can now be
computed in the effective low--energy theory. At this point characteristic
differences between the different spin projections of the operator
emerge. A parametrically large contribution can only come from contractions
of the four--fermionic operator, Eq.(\ref{O_eff}), with closed quark loops,
in which the quark momenta run up to the ultraviolet cutoff,
$\bar\rho^{-1}$ (see Ref.\cite{BPW97} for details).  One finds that in the
case of the twist--4 matrix element, $f^{(2)}$, Eq.(\ref{f2_def}), the loop
is $\propto \bar\rho^{-2}$ (``quadratically divergent''), while in the case
of the twist--3 matrix element, $d^{(2)}$, Eq.(\ref{d2_def}), it is only of
order $\log M\bar\rho$, (``logarithmically divergent''). As a result, we
obtain a parametric difference of the twist--4 and 3 matrix elements:
\[
\begin{array}{lrclcl}
\mbox{twist 4:} & f^{(2)}
&\sim& (M\bar\rho )^0 &\sim& 
{\displaystyle \left(\frac{\bar\rho}{R}\right)^0 } , \\[.3cm]
\mbox{twist 3:} & d^{(2)} &\sim& (M\bar\rho )^2 \log M\bar\rho 
&\sim& {\displaystyle
\left(\frac{\bar\rho}{R}\right)^4 \log \left(\frac{\bar\rho}{R}\right) .}
\\
\end{array}
\]
This remarkable result can be traced back to two circumstances: the $O(4)$
symmetry of the field of a single (anti--) instanton, and the diluteness of
the instanton medium.
\par
The parametric difference between the twist--3 and 4 matrix elements also
reflects itself in the numerical values. For the flavor non-singlet nucleon
matrix element $f^{(2)}_{\mbox{\scriptsize NS}}$ (the NS part is leading in
the $1/N_c$--expansion) we obtain a value of $-0.1$, while
$d^{(2)}_{\mbox{\scriptsize NS}}$ was estimated to be of the order of
$10^{-3}$.\footnote{An accurate calculation of $d^{(2)}$ in this approach
would require to take into account two--instanton contributions in the
effective operator; this calculation is presently being done. The
order--of--magnitude estimate in Ref.\cite{BPW97} is based only on the
one--instanton contribution.} Thus, the instanton vacuum implies a clear
numerical ordering of higher--twist matrix elements: 
$|d^{(2)}| \ll |f^{(2)}|$. Again we stress that this ordering is not
accidental, but a direct consequence of qualitative properties of the
instanton medium.  This is at variance with the QCD sum rules estimates,
and also of bag model calculations, which typically give values of
$f^{(2)}_{\mbox{\scriptsize NS}}$ and $d^{(2)}_{\mbox{\scriptsize NS}}$ of
the same order, see Table 1.  (For a critical discussion of the QCD sum
rule calculations see Ref.\cite{Ioffe}.) Note also that the instanton
result for $f^{(2)}_{\mbox{\scriptsize NS}}$ agrees in sign with the QCD
sum rule results, but not with the bag model.
\par
It is interesting that the small value of $d^{(2)}$ suggested by the
instanton vacuum seems to be in agreement with recent measurements of $g_2$
by the E143 collaboration. Ref.\cite{Rock98} reports a value of
$d_{\mbox{\scriptsize NS}}^{(2)} \equiv 3 (d_p^{(2)} - d_n^{(2)}) = (2.4
\pm 78) \times 10^{-3}$, which agrees well with the order--of--magnitude
estimate of Ref.\cite{BPW97} (note that $d_p^{(2)}$ and $d_n^{(2)}$
individually are of comparable magnitude, see Ref.\cite{Rock98}).  This value
is much smaller than our estimate for $f^{(2)}_{\mbox{\scriptsize NS}}$,
which is of order $10^{-1}$.  Unfortunately, the present data are not
sufficient to unambiguously extract $f^{(2)}$ from power corrections with
sufficient accuracy in order to make a quantitative comparison of $f^{(2)}$
and $d^{(2)}$ \cite{JM97}. It would be interesting to see if by {\it
imposing} the condition $|d^{(2)}| \ll |f^{(2)}|$ from the start one could
extract $f^{(2)}$ more accurately.
\par
To summarize, we have shown that the instanton vacuum with its inherent
small parameter, $\bar\rho / \bar R$, implies a parametrical (and
numerical) hierarchy of the spin--dependent twist--3 and 4 matrix elements:
$|d^{(2)}| \ll |f^{(2)}|$.  This qualitative prediction can be confronted
with structure function data. Extending this analysis further ({\it e.g.}\
to higher moments and to other structure functions) one could use the rich
information available from DIS to learn about properties of
non-perturbative vacuum fluctuations.
%
%
\noindent
\begin{table}[t]
\begin{center}
\begin{tabular}{|l|l|l|c|}
\hline
& $f^{(2)}_{\mbox{\scriptsize NS}} \times 10^1$ 
& $d^{(2)}_{\mbox{\scriptsize NS}} \times 10^1$ & 
$\mbox{scale}/{\rm GeV}^2$ \\ \hline
Instanton vacuum \cite{BPW97} & $-1.0$  & $\sim 0.01$ & $0.4$ 
\\
QCD sum rules \cite{BBK90} &  $-2.0$  & $\;\;\;\, 0.7$  &  1   
\\
QCD sum rules \cite{Stein95} & $-0.7$ & $\;\;\;\, 0.7$  & 1   
\\
Bag model \cite{JiU94} &  $\;\;\; 1.1$  &        $\;\;\;\, 0.6$  & 5 
\\
E143 \cite{Rock98} & & $\;\;\;\, 0.024 \pm 0.78 $ & 5  
\\
\hline
\end{tabular}
\end{center}
\caption[]{Numerical results for the flavor--nonsinglet spin--dependent
twist--4 and 3 matrix elements 
$f^{(2)}_{\mbox{\scriptsize NS}} \; \equiv 3 (f^{(2)}_p - f^{(2)}_n )$ and
$d^{(2)}_{\mbox{\scriptsize NS}}$, Eqs.(\ref{d2_def}) and (\ref{f2_def}).
The instanton result for $d^{(2)}_{\mbox{\scriptsize NS}}$ is an
order--of--magnitude estimate (see the text).}
\label{table_numeric}
\end{table}
\par
It is a pleasure to thank D.I.\ Diakonov, V.Yu.\ Petrov, P.V.\ Pobylitsa
for many interesting discussions, and K. Goeke for encouragement and
support.
\end{document}